%% file: main.tex
\documentclass[a4paper,11pt]{article}
\usepackage{jinstpub} 
\usepackage{lineno}
\usepackage{hyperref}

\title{\boldmath High-rate, high-resolution single photon X-ray imaging: Medipix4, a large 4-side buttable pixel readout chip with high granularity and spectroscopic capabilities}

\author{V. Sriskaran$^a$, J. Alozy$^a$, R. Ballabriga$^a$, M. Campbell$^a$, P. Christodoulou$^{a,b}$, E. Heijne$^b$, A. Koukab$^c$, T. Kugathasan$^{a^{**}}$, X. Llopart$^a$, M. Piller$^{a,d}$, A. Pulli$^a$, J-M Sallese$^c$, L. Tlustos$^{a,b}$
}
\affiliation{$^a$CERN, EP Department, 1211 Geneva, Switzerland}
\affiliation{$^b$IEAP Institute of Experimental and Applies Physics,\\
                Czech Technical University in Prague, Husova 240/5, 110 000 Prague 1, Czech Republic}
\affiliation{$^c$EPFL, 1015 Lausanne, Switzerland}
\affiliation{$^d$ Graz University of Technology, Institute of Electronics, 8010 Graz, Austria}
\affiliation{$^{**}$ now with University of Geneva, Geneva, Switzerland}

\emailAdd{michael.campbell@cern.ch}

\abstract{The Medipix4 chip is the latest member in the Medipix/Timepix family of hybrid pixel detector chips aimed at high-rate spectroscopic X-ray imaging using high-Z materials. It can be tiled on all 4 sides making it ideal for constructing large-area detectors with minimal dead area. The chip is designed to read out a sensor of 320 x 320 pixels with dimensions of 75 µm x 75 µm or 160 x 160 pixels with dimensions of 150 µm x 150 µm. The readout architecture features energy binning of the single photons, which includes charge sharing correction for hits with energy spread over adjacent pixels. This paper presents the specifications, architecture, and circuit implementation of the chip, along with the first electrical measurements.}


\begin{document}
\maketitle
\flushbottom

\input{Introduction}

\input{chip_description}

\input{pixel_cell}
\input{Results}
\input{Summary}


\acknowledgments

This work was carried out in the context of the Medipix4 Collaboration. The first results of electrical characterization were presented at 24th International Workshop On Radiation Imaging Detectors (Oslo, June 2023). The authors would like to express their sincere gratitude to the members of the Collaboration for their valuable contributions to the formulation of the specifications, as well as their financial support. We would also like to thank Bas van der Heijden and Vincent at Nikhef for their help in modifying the firmware on the SPIDR4 readout system.





\bibliography{biblio}

\end{document}

%% file: Introduction.tex
\section{Introduction}
\label{sec:intro}

Over the past two decades, 4 successive Medipix collaborations has been established. These collaborations aim to exploit the knowledge acquired from advancements in high-energy physics to develop cutting-edge hybrid pixel detectors, enabling the precise detection of individual X-ray photons or particles on a per-event basis~\cite{HEIJNE_HIST}. These technologies have diverse applications across scientific domains, including medical imaging, X-ray cameras at synchrotrons, X-ray-based material analysis, electron microscopy, among others.

First, the Medipix1 chip demonstrated the principle of a single photon counting architecture within a pixel pitch of 170 µm, and showcased the feasibility of X-ray imaging without noise hits by using a pulse processing front end while setting the detection threshold well above the level of background noise \cite{MPX1_campbell}. Medipix2 proved the feasibility of spectroscopic imaging with a compact pixel pitch of 55 µm by using  dual thresholds per pixel \cite{Medipix2}. However, the reduced pixel dimensions led to significant charge sharing between pixels due to diffusion during charge collection and fluorescence photons in high-Z materials \cite{MPX2_chargesharing,Tlustos_thesis}. The readout electronics underwent a transition from single photon counting to a single photon processing architecture with the introduction of Medipix3RX. A novel scheme implementing an inter-pixel algorithm directly on the 55 µm pixel eliminated the energy spectral distortion produced by charge diffusion~\cite{MPX_TPX_chips,MPX3_prototype}. Medipix3RX also introduced the option of connecting one pixel in 4 to a sensor with 110 µm pixel pitch.

Nonetheless, the Medipix3RX detectors can be abutted only on three sides, as one side of the chip is reserved for control logic and IO. This complicates the realization of a continuous large-area detector. The Medipix4 presented in this paper, follows the advancements of the Timepix4 chip and enables the application-specific integrated circuit (ASIC) to be tilled along all four sides with minimal dead-area~\cite{Llopart_2022}. Another constraint in medical X-ray computed tomography (CT) and X-ray imaging arises from pulse pile-up, which is attributed to the inherent dead time of counting systems~\cite{20_years_CT}. Some recent photon counting detectors have begun developing on-pixel schemes to compensate for this effect and to increase the count-rate performance when employing monochromatic sources~\cite{PILATUS3,Eiger2,SPHIRD}. In this latest Medipix iteration, a new pulse processing mechanism is integrated directly into the pixel to filter pile-up events that can occur with polychromatic radiation. Medipix4 focuses on improving the spectral fidelity using both  charge sharing correction and pile-up filtering method.

Furthermore, the majority of current X-ray imaging systems are tailored for particular applications, such as employing small pixel pitches in synchrotron applications to achieve high count-rate capability while sacrificing spectral fidelity due to charge sharing and fluorescence effects. Conversely, larger pixel sizes are employed in most CT imaging systems to cope more with fluorescence in high-Z materials, while compromising spatial resolution (complete review on existing photon counting detectors for X-ray imaging and its limiting factors can be found in our paper~\cite{Rafa_CT}). The ASIC presented in this paper is highly programmable to accommodate a large range of applications. The Medipix4 chip has been designed to facilitate large-scale applications, enabling high-rate spectroscopic X-ray imaging at fine pitch using high-Z materials while maintaining spectral accuracy through the implementation of an inter-pixel architecture to correct the impact of charge sharing. The structure of this paper is as follows: Section~\ref{sec:chip_descr} starts by outlining the key specifications of the Medipix4 chip alongside its floorplan. In Section~\ref{sec:pixel_arch}, the architecture of the analog and digital pixel cells is described. Section~\ref{sec:results} focuses on the initial results of electrical characterization. We conclude with a summary and an outline of future plans.

%% file: chip_description.tex
\section{Medipix4 chip floorplan and specifications}
\label{sec:chip_descr}

\begin{figure}[htbp]
\centering
\includegraphics[width=1\textwidth]{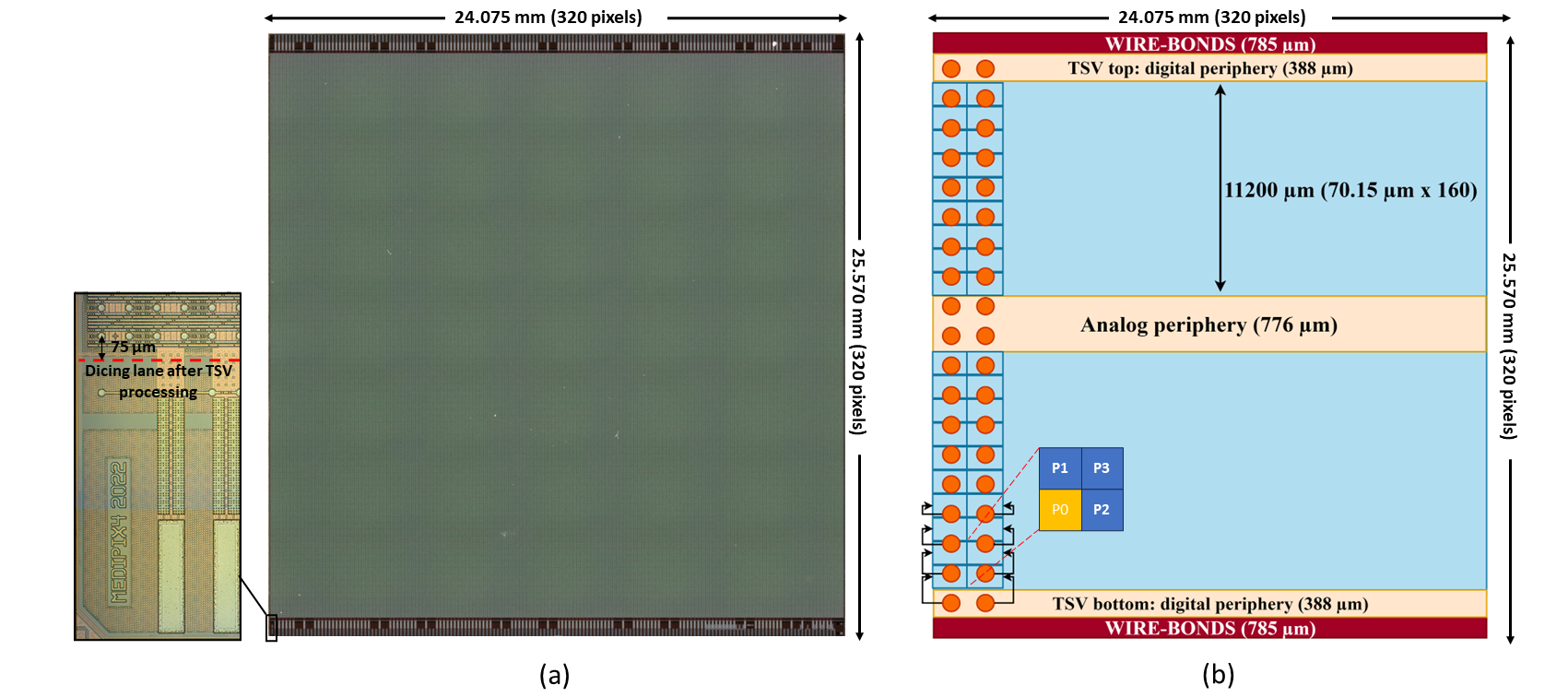}
\caption{(a) The picture of the Medipix4 chip is presented, with a magnified inset located in the bottom corner. This inset shows the wire bond extenders that can be diced off after TSV processing. (b) The floorplan of the chip is shown schematically (not to scale). The readout pixels within the blue region measure 75 µm x 70.15 µm and are connected to an array of bump bonding pads with 75 µm pitch in both directions. This allows space to accommodate the peripheral circuits at the top, middle, and bottom of the chip.     \label{fig:i}}
\end{figure}

The ASIC measures 24.075 mm x 25.570 mm and has been designed using an 8-metal layer commercial 130 nm technology process and has a power supply voltage of 1.2 V. The chip is designed to read out a sensor of 320 x 320 pixels with dimensions of 75 µm x 75 µm (\textit{fine pitch mode}) or 160 x 160 pixels with dimensions of 150 µm x 150 µm (\textit{spectroscopic mode}).  The top surface of the Medipix4 ASIC is fully covered with a matrix of 320 x 320 bump bonding pads (with 20 µm octagonal opening) distributed on a 75 µm square pitch. The Medipix4 chip adopts the Timepix4 layout configuration, thereby attaining a complete sensitive area and enabling tiling along its four sides~\cite{Llopart_2022}. The readout pixels and peripheral circuits are positioned beneath the uniformly distributed bump bonding pads. This arrangement results in the readout pixels being smaller than the sensor pixels in one direction to incorporate the peripheral circuits. The readout electronics consist of two arrays, each containing 320 x 160 pixels with dimensions of 75 µm x 70.15 µm. A redistribution layer (RDL) is integrated using the top metal layers to establish connections between the pads and their corresponding readout electronics. As this approach involves a non-intimate connection between the readout and sensor pixels, it comes at the 'cost' of an additional 80 fF to the input capacitance of the front-end electronics. The primary challenge revolves around ensuring uniformity in input capacitance and providing adequate shielding to suppress any coupling with neighboring pixels and the chip's readout electronics~\cite{Sriskaran_thesis}. The chip incorporates three peripheral regions, a space of 776 µm between the two matrices of readout pixels forming the analog periphery and 388 µm on each side for the digital peripheries. The analog periphery is composed of the blocks required for biasing the analog circuits within the pixels. The digital peripheries include control logic, data output SLVS (programmable between 1 to 8 per edge periphery), and the through-silicon via (TSV) structures. The chip offers connectivity to a printed circuit board through either standard wire bonding or via TSV technology. Wire-bond extenders are situated at the upper and lower edges of the ASIC, enabling wafer-level probing and single ASIC wire bonding without the necessity of TSV processing (at the expense of 93.5${\%}$ of active area). The wire-bond extenders can be removed by dicing when input/output interconnections are established with TSV processing. The dicing is performed along the red dashed line depicted in the magnified inset on figure~\ref{fig:i} and located 75 µm from the first bump-pad (and symmetrically from the last). In the TSV scenario, the die size is reduced to 24.075 mm x 24.075 mm, covering 99.37${\%}$ of the active area without compromising chip functionality.

The Medipix4 chip is designed to operate in two primary acquisition modes: \textit{single pixel mode} (SPM) and \textit{charge summing mode} (CSM). In SPM, each pixel operates like conventional single-photon counting architecture, operating independently from its neighboring pixels. Conversely, in CSM, the inter-pixel architecture reconstructs the total charge within overlapping clusters of 2 x 2 pixels, guided by an arbitration circuit that allocates the hit to the pixel with the highest charge deposition. The complete reconstructed charge is then attributed to the winning pixel within a neighbourhood of 9 pixels. This charge processing architecture corrects the spectral distortion that arises from charge diffusion across the sensor material~\cite{MPX3_prototype,MPX3,Koenig}. 

The Medipix4 architecture proposes pixel size programmability with a 75 µm pixel pitch in \textit{fine pitch mode} (FPM) and 150 µm in \textit{spectroscopic mode} (SM). FPM is particularly well-suited for applications that utilize Si or GaAs as sensor materials~\cite{Koenig_GaAs}, while SM is better suited for high-Z materials like CdTe, CdZnTe, or perovskite~\cite{Koenig}. Please note that using the fine pitch pixel with high-Z materials and with no charge sharing correction could lead to multiple hit counts for a single incoming photon due to the potential deposition of fluorescence photons far from the original impact point. For instance, CdTe has more than an 80\% chance of generating fluorescence photons, and the mean free path of those photons is comparable to the fine pixel size: 58 µm for Te and 111 µm for Cd~\cite{Pennicard_CSM}. Similar to Medipix3RX, both SPM and CSM are available for both sensor pixel pitches, providing four different charge collection areas. Medipix3RX with a pixel pitch of 110 µm still exhibited some escape peaks from fluorescence in CdTe, even with charge-sharing correction (providing a charge collection area of 220 µm x 220 µm)~\cite{Koenig_2012_CdTe}. For that purpose, a 75 µm pixel pitch was selected for the Medipix4 readout pixel, enabling a larger collection area of 300 µm x 300 µm in the SM-CSM configuration.

%% file: pixel_cell.tex
\section{Medipix4 pixel architecture and readout}
\label{sec:pixel_arch}
\subsection{Pixel schematic and analog processing modes}
\begin{figure}[htbp]
\centering
\includegraphics[width=1\textwidth]{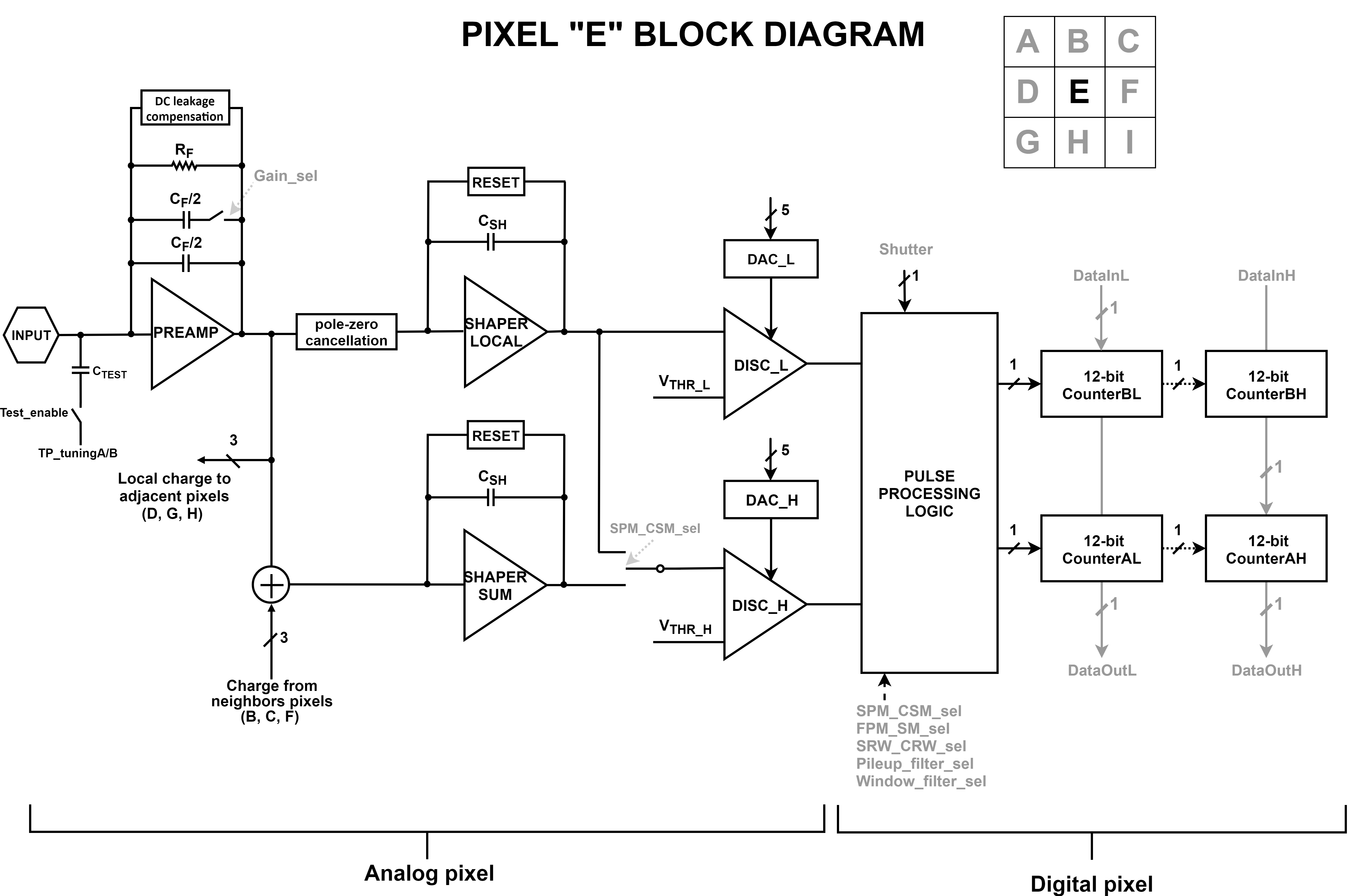}
\caption{Simplified block diagram of the Medipix4 pixel cell. The communication of the pixel E with its adjacent pixels is shown in \textit{fine pitch mode}.     \label{fig:f2}}
\end{figure}

The block diagram of the Medipix4 pixel cell is illustrated in Figure~\ref{fig:f2}. In the analog circuitry, the initial stage contains a charge sensitive amplifier (CSA) with a programmable feedback capacitance. This feedback capacitance can be configured between 5 fF, achieving the highest gain in the \textit{low noise mode} (LNM), and 10 fF in the \textit{ultra-fast mode} (UFM). Unlike its predecessors, the Medipix4 front-end is only sensitive to negative charges originating from the sensor (electron collection). A test injection capacitance of 5 fF allows for electrical stimuli up to 25 ke$^{-}$. The chip provides access to two separate test pulses, enabling the injection of different charges into neighboring pixels. To manage the DC leakage current from the sensor, the first stage incorporates a compensation network with the ability to accommodate currents up to 50 nA per pixel. A pole-zero cancellation circuit follows the CSA. Subsequently, the signal is directed to a pulse-shaping amplifier that offers adjustable peaking and discharge times. The new shaper amplifier demonstrated in simulation reduced baseline drift at high flux when compared to the Krummenacher feedback topology employed in previous Medipix and Timepix iterations~\cite{KRUMMENACHER1991527}. This enhancement is attributed to the filter circuitry within the DC compensation network, featuring back-to-back connected transistors that provide a significant equivalent resistance while using small silicon area~\cite{BtB_MOSFET,Thanu_BH}.

In SPM, the shaper denoted as "LOCAL" is connected to two parallel comparators, each equipped with its own distinct energy threshold, which is independently adjustable. To address threshold disparities from one pixel to another, two 5-bit tuning digital-to-analog converters (DAC) are connected to the comparators, effectively compensating for any mismatches. This mode enables simultaneous imaging with dual thresholds.

The charge sharing correction is performed in CSM by enabling the summing shaper circuit called "SUM." This circuit reconstructs the charge within overlapping clusters of 2 x 2 pixels, thereby establishing an equivalent collection area of 150 µm x 150 µm. Each pixel sums its own contribution, as well as the contributions from pixels in its east, north, and north-east locations. In this summing mode, the noise from the four pixels is combined in quadrature, resulting in a twofold increase in the total noise. For instance, pixel E forwards its local charge to the summing nodes of pixels D, G, and H while receiving the charge contribution from pixels B, F, and C. Concurrently, a network of arbitration circuits within the digital pixel determines the pixel with the highest charge deposition among its neighbors, providing an overall 75 µm spatial resolution. This is possible using the time-over-threshold information. These steps guarantee the full reconstruction of the charge deposited in 2 x 2 pixels through an analog scheme, while the digital scheme allocates the reconstructed charge to a single pixel, preventing multiple counting for a single hit. This analog charge summing overcomes the limitation of sub-threshold charge loss that occurs when implementing digital charge reconstruction \cite{TLUSTOS_Subthr_loss}. There is one discriminator ($DISC\_L$) with an energy threshold placed above the noise floor for the arbitration of the largest charge in the local area, and a different threshold for the reconstructed charge ($DISC\_H$)~\cite{SRISKARAN_MPX4}.

When programming the Medipix4 in \textit{spectroscopic mode} (SM), the readout pixels are organized into a cluster of four pixels (labeled as P0, P1, P2, and P3 in Figure~\ref{fig:i}). Among these, only the readout pixel P0, referred to as the "master" pixel, is bump bonded to the 150 µm pitch sensor pixel. The front-end circuits within the remaining three pixels ("slave" pixels) are deactivated, while their energy discriminators and counters are linked to the "master" pixel. In the scenario involving a 150 µm pixel in \textit{single pixel mode}, a total of eight independent thresholds are accessible. However, when applying the charge-sharing algorithm with the 150 µm pixel, one energy threshold is allocated to the arbitration network denoted SPM\_A, while seven thresholds are dedicated to the reconstructed charge. An overview of the different modes of operation and thresholds is provided in Table \ref{tab:operation_modes}.
\begin{table}[ht]
\centering
\small
\begin{tabular}{|c|c|c|c|}
\hline
\textbf{Operation modes} & \textbf{Spatial resolution}  & \textbf{Charge collection area} & \textbf{Thresholds} \\
\hline
FPM-SPM & 75 µm & 75 µm x 75 µm & 2 \\
\
FPM-CSM & 75 µm & 150 µm x 150 µm & 1 $SPM_{A}$ + 1  CSM \\
\hline
SM-SPM & 150 µm & 150 µm x 150 µm & 8 \\
\
SM-CSM & 150 µm & 300 µm x 300 µm & 1 $SPM_{A}$ + 7 CSM\\
\hline
\end{tabular}
\caption{Summary of operation modes in Medipix4.}
\label{tab:operation_modes}
\end{table}

The analog pixel is highly configurable to accommodate a large range of different applications. \textit{Ultra-fast mode} (UFM), for instance, may be advantageous for applications such as computed tomography and synchrotron cameras, that require a very high count-rate capability. We simulated a deadtime not exceeding 130 ns (at 13.7 $ke^-$ or 60 keV in CdTe). This will enable a count-rate capability up to 19 x $10^6 photons.mm^{-2}.s^{-1}$ at 10\%  hit loss, all while maintaining a 150 µm pixel pitch and remaining unaffected by charge sharing and fluorescence effects. This fast photon processing comes at the expense of a higher electronic noise. The latter can be improved in \textit{low noise mode} (LNM) for a better energy resolution. Across all operational modes, the power consumption per unit area is intentionally maintained well below 0.5 $W/cm^{-2}$. This design choice allows the chip to operate without the need for active cooling~\cite{Rafa_CT}. The power consumption is reduced in the 150 µm pixel configuration, wherein the front-end circuits of the slave pixel are deactivated.

\subsection{Digital processing modes}

Medipix4 facilitates spectral information extraction through analog pulse height analysis using two to eight energy thresholds. In general, window discrimination is used in order to optimize the counter depths, avoiding counter saturation in low energy bins seen in Medipix3RX. 
\begin{figure}[htbp]
\centering
\includegraphics[width=1\textwidth]{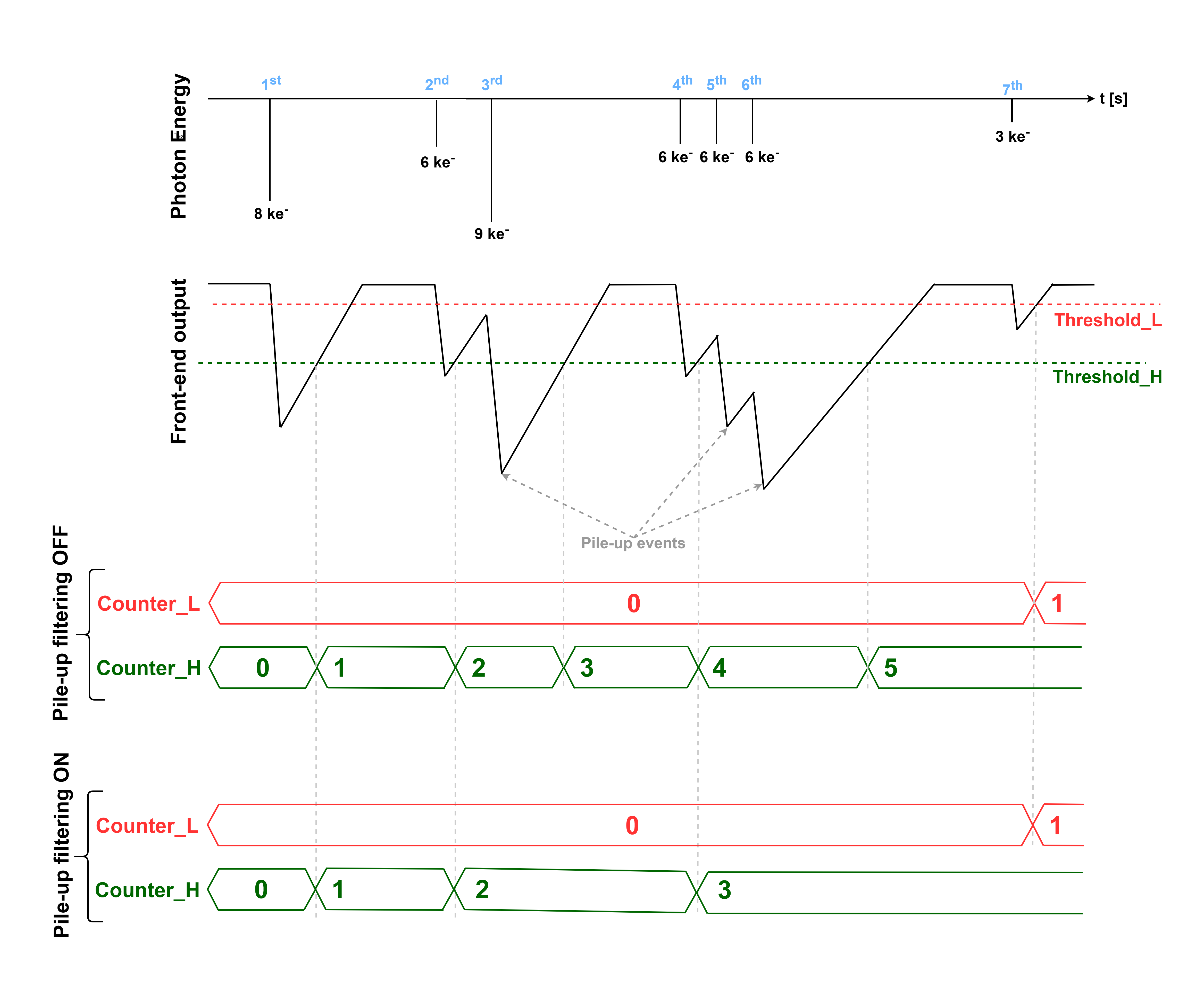}
\caption{Working principle of the digital pixel when the chip is configured in \textit{fine pitch mode} with \textit{single pixel mode} and \textit{pile-up filtering mode} enabled (and disabled for comparison) while keeping \textit{window discrimination mode} on.    \label{fig:pileup}}
\end{figure}

Medipix4 is designed to provide fast CT imaging capability while maintaining spectral fidelity. In addition to the charge correction algorithm, a new pile-up filtering mechanism is integrated into the digital pixel to mitigate the impact of pulse pile-up on the energy spectrum at the expense of some lost hits. Pulse pile-up occurs when the analog front-end does not return to the baseline between two consecutive events, leading to an incorrect assignment of subsequent events to their energy bins. Consequently, the system's reliability diminishes at high flux due to poor spectral fidelity in measurements. The Medipix4 digital pixel can be configured in \textit{pile-up filtering mode} to discard the events arriving during the discharge time of previous events. Figure~\ref{fig:pileup} illustrates the pile-up filtering mechanism when the digital pixel is configured to operate in \textit{fine pitch mode} and \textit{single pixel mode} with \textit{window discrimination} and \textit{pile-up filtering}, using two thresholds (the lower energy threshold associated to $Counter\_L$, placed close to the noise floor, and the higher threshold to $Counter\_H$). Thanks to window discrimination, the first event is only allocated to the higher energy bin. The filtering mechanism discards events 3, 5, and 6 as the analog signal did not return to its baseline (which would have triggered the lowest energy threshold). This mechanism does not detect peak pile-up events where coincidences occur around the initial event's peaking time. Only events falling on its tail are detected by the digital pixel. Detailed measurements on the pile-up filtering mechanism and the count-rate capability of our ASIC when bonded to a sensor will be published when they become available.

\subsection{Pixel matrix readout}

Each pixel contains four 12-bit shift registers that function as counters during acquisition or as shift registers for reading out the collected data. The pixel readout can be programmed in \textit{sequential read/write} (SRW) or \textit{continuous read/write} (CRW) modes. In SRW mode, events are stored in counters that can be configured in a 1-bit, 2-bits, 12-bits, or 24-bits depth format. When the \textit{shutter} signal is low, the counts of the energy bins are transmitted to the end of the chip column (top and/or bottom edge periphery) for subsequent readout. CRW mode enables readout dead-time-free operation for all energy thresholds in both \textit{fine pitch mode} and \textit{spectroscopic mode}. In this last scenario, pulses can be recorded in either a 1-bit or 12-bit depth format, while the readout is executed through the other 1-bit or 12-bit depth counter.

\begin{table}[htbp]
\centering
\small
\begin{tabular}{|c|c|c|c|c|}
\hline
\multicolumn{1}{|c|}{} & \multicolumn{2}{c|}{\textbf{Fine Pitch Mode (75 µm pixel)}}& \multicolumn{2}{c|}{\textbf{Spectroscopic Mode (150 µm pixel)}}\\
\multicolumn{1}{|c|}{} & \multicolumn{1}{c|}{SPM} & \multicolumn{1}{c|}{CSM} & \multicolumn{1}{c|}{SPM} & \multicolumn{1}{c|}{CSM} \\
\hline
\multicolumn{1}{|c|}{On-pixel thresholds} & \multicolumn{1}{c|}{2} & \multicolumn{1}{c|}{1 $SPM_A$ + 1 CSM } & \multicolumn{1}{c|}{8} & \multicolumn{1}{c|}{1 $SPM_A$ + 7 CSM} \\
\hline
\multicolumn{1}{|c|}{Counter depth Sequential} & \multicolumn{2}{c|}{2 x [1, 2, 12, or 24] bits} & \multicolumn{2}{|c|}{8 x [1, 2, 12, or 24] bits} \\
\multicolumn{1}{|c|}{Read/Write [dynamic Range]} & \multicolumn{2}{c|}{} & \multicolumn{2}{c|}{} \\
\hline
\multicolumn{1}{|c|}{Counter depth Continuous} & \multicolumn{2}{c|}{2 x [1 or 12] bits} & \multicolumn{2}{|c|}{8 x [1 or 12] bits} \\
\multicolumn{1}{|c|}{Read/Write [dynamic Range]} & \multicolumn{2}{c|}{} & \multicolumn{2}{c|}{} \\
\hline
\multicolumn{1}{|c|}{On-pixel digital modes} & \multicolumn{4}{c|}{Window discrimination threshold mode
} \\
\multicolumn{1}{|c|}{} & \multicolumn{4}{c|}{Pile-up filtering mode} \\
\hline
\multicolumn{1}{|c|}{Readout direction modes} & \multicolumn{4}{c|}{Split readout for independent readout from both sides} \\
\multicolumn{1}{|c|}{} & \multicolumn{4}{c|}{Conveyor belt readout (towards top or bottom periphery)} \\
\hline
\multicolumn{1}{|c|}{Output ports} & \multicolumn{4}{c|}{ 1 to 16 SLVS (up to 640 Mbps per port)}\\
\hline
\end{tabular}
\caption{Overview of readout modes in Medipix4 ASIC.\label{tab:Readout_mode}}
\end{table}

 The Medipix4 pixel matrix offers two distinct readout modes: \textit{conveyor readout mode} and \textit{split readout mode}. \textit{Conveyor mode} is specifically designed for scenarios necessitating a continuous scanning, like for imaging moving objects~\cite{Conveyor_imaging_Dudak} or for X-ray laminography. In this configuration, the entire pixel matrix is read out via a single edge periphery utilizing 1, 2, 4, or 8 output serializers. Alternatively, in the \textit{split readout mode}, both edge peripheries independently read out their respective 320 x 160 matrices through a maximum of 16 SLVS links. A full overview of the different readout modes is provided in table \ref{tab:Readout_mode}.

The chip's maximal frame rate is contingent upon a number of parameters, including the counter bit depth (which must be chosen in accordance with the exposure time to prevent counter overflow), the pixel readout mode, the number of parallel output links (ranging from 1 to 16), and the frequency of the data acquisition clock (up to 320 MHz). For instance, the use of \textit{split readout mode} results in a maximum frame rate that is twice as high as that achieved through the \textit{conveyor readout mode}. Furthermore, higher frame rates can be attained by selectively reading a sub-area within the pixel matrix. This is done by specifying the number of rows in the matrix to be read out. For example, in a large multi-module detector, this feature may emulate a smaller-size detector with significantly elevated frame rates. Table \ref{tab:frame_rate} provides the frame rate capability of the chip in a few selected readout configurations. For instance, when configuring the chip to continuously read out a single row in 12-bit format, a frame rate of 1.176 MHz is achieved, compared to 8.32 kHz obtained by reading the full matrix. Alongside this readout feature, the number of frames and the inter-frame wait times can be programmed.

\begin{table}[htbp]
\centering
\small
\begin{tabular}{|c|c|c|c|c|}
\hline
\textbf{Counter length} & \textbf{Readout mode}  & \textbf{Full matrix readout} & \textbf{Links} & \textbf{Frame rate [fps]}\\
\hline
1 & Split readout \& CRW & Yes & 16 & 99 K\\
2 & Split readout \& SRW & Yes & 16 & 49.5 K\\
12 & Split readout \& CRW & Yes & 16 & 8.32 K\\
24 & Split readout \& SRW & Yes & 16 & 4.16 K\\
\hline
1 & Conveyor readout \& CRW & Yes & 8 & 49.7 K\\
2 & Conveyor readout \& SRW & Yes & 8 & 24.87 K\\
12 & Conveyor readout \& CRW & Yes & 8 & 4.16 K\\
24 & Conveyor readout \& SRW & Yes & 8 & 2.08 K\\
\hline
1 & Split readout \& CRW & No, 1 row & 8 & 6154 K\\
12 & Split readout \& CRW & No, 1 row & 8 & 1176 K\\
\hline
\end{tabular}
\caption{Maximum frame-rate capability of Medipix4 for full matrix readout and partial readout.}
\label{tab:frame_rate}
\end{table}

%% file: Results.tex
\section{Electrical measurements}
\label{sec:results}

\begin{figure}[htbp]
\centering
\includegraphics[width=1\textwidth]{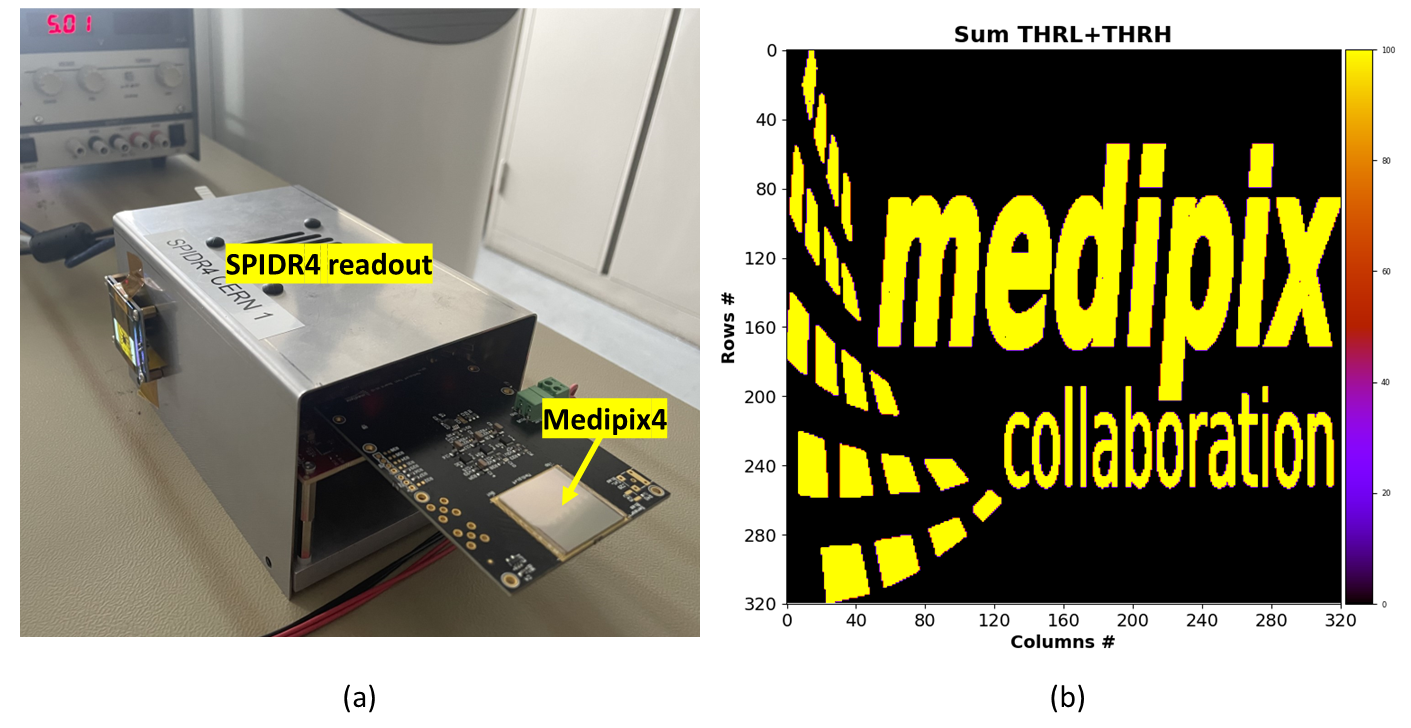}
\caption{(a) shows the Medipix4 chip mounted on a printed board circuit using the wire bond extenders option and read out using the SPIDR4 system. (b) The full chip imaging using test-pulses when the pixel is configured in FPM-SPM and with two thresholds. 21 pixels were masked during this measurement.     \label{fig:setup}}
\end{figure}

A single Medipix4 chip without sensor was mounted on a printed circuit board using the wire bonding option. To enable initial electrical characterization, slight adaptations were made to the readout system SPIDR4, originally designed for Timepix4 \cite{TMPX4_timing}, with valuable assistance from Nikhef. Access to the chip via slow control allowed for the first round of electrical tests. The test setup is shown on the left hand side in Figure \ref{fig:setup}. Furthermore, dedicated efforts towards developing a specialized readout system for Medipix4 are underway. The outcomes presented in this paper concern the second iteration of Medipix4. In the first Medipix4 version, within the analog front-end pixel, the models exhibited an underestimation of the substrate leakage current in the thick oxide transistors employed as pseudo-resistors. Consequently, certain pixels within the matrix failed to lock onto the correct reference level, even after pixel threshold equalization. With the updated current version, the sensitivity to substrate leakage has been addressed.
 The right plot in Figure \ref{fig:setup} depicts the image of the Medipix4 collaboration logo obtained by sending test pulses to the pixel matrix and reading out the accumulated hits in the two counter bins. In this case, the chip was configured in \textit{fine pitch mode} and \textit{single pixel mode}. Only 21 pixels are masked, demonstrating a good yield after threshold equalization.

\begin{figure}[htbp]
\centering
\includegraphics[width=1\textwidth]{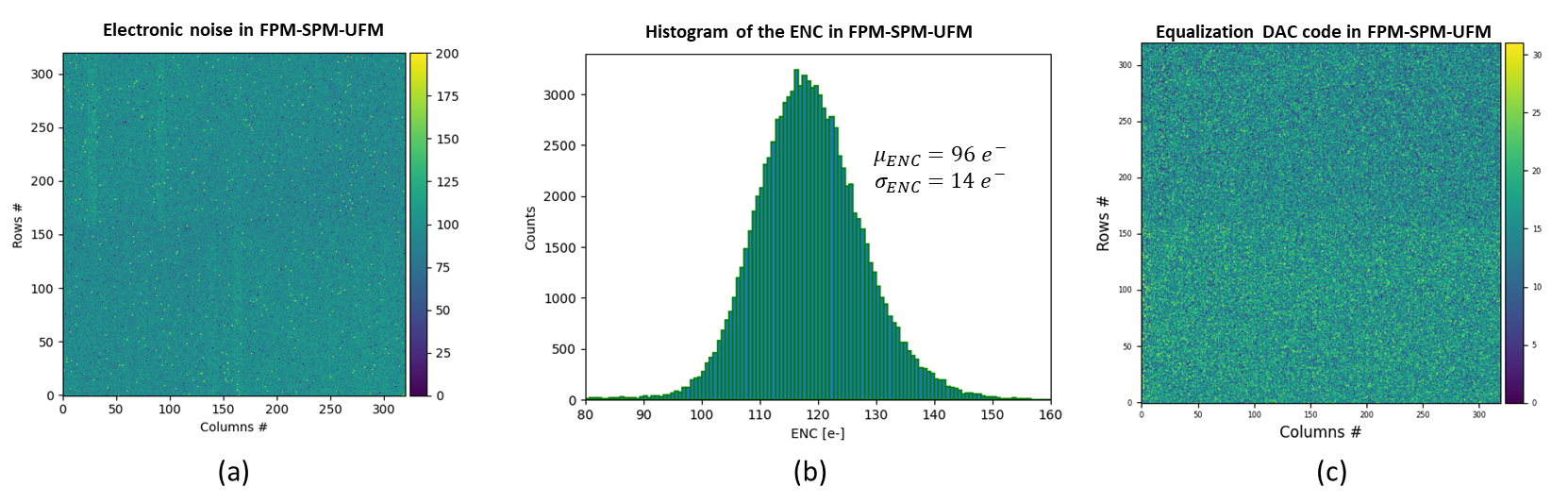}
\caption{The measured electronic noise in r.m.s electrons when the chip is programmed in FPM-SPM for which the analog front-end configured in \textit{ultra-fast mode} (a) and the histogram (b). (c) The pixel DAC code for equalization exhibits a minor systematic offset between the top and bottom matrices, arising from a slight deviation in the pixel reference definition for the two matrices. It is important to highlight that, despite this marginal offset, the threshold dispersion following pixel equalization consistently maintains a magnitude smaller than the electronic noise, as illustrated in Table \ref{sec:results}.   \label{fig:noise}}
\end{figure}

 Figure \ref{fig:noise} presents the measured electronic noise when configuring the analog front-end in \textit{ultra-fast mode}. The measured noise matches with the expected values, with a mean noise level of 116 $e^-$ and a dispersion of 10 $e^-$ r.m.s. Please note that that these measurements are performed on a bare readout chip using the electrical test pulses. The precision of these measurements relies on knowledge of the test pulse capacitance, which carries a tolerance of approximately $\pm 30\%$. For a more precise assessment of the front-end parameters, a subsequent measurement will be undertaken using radioactive sources following the flip chip bonding of the readout chip to a sensor.

\begin{figure}[htbp]
\centering
\includegraphics[width=1\textwidth]{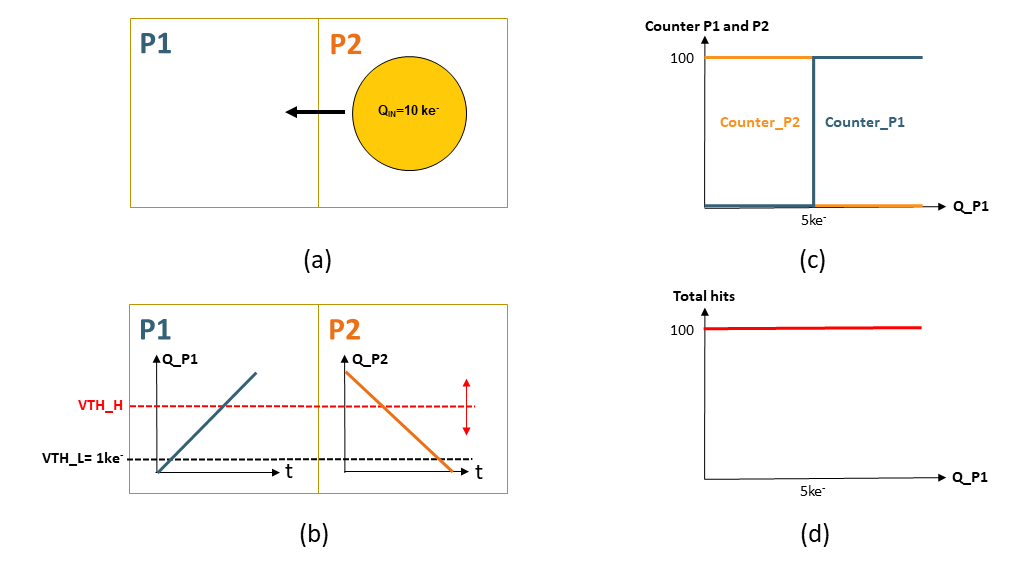}
\caption{(a) depicts a scenario wherein a 10 $ke^-$ input charge is moved from the right sensor pixel P2 to the left sensor pixel P1. (b) displays the charge distribution at the front-end input for both pixels, resulting from charge sharing at the pixel edges. (c) illustrates the ideal pixel response, where the counter associated with the higher energy threshold (VTH\_H) is capable of accurately reconstructing both the energy and spatial characteristics of incoming hits. It is important to note that this holds true when the higher energy threshold is set lower than the input energy. (d) shows the total hits registered by both pixels in an ideal case, highlighting the absence of hit loss or multiple counts.     \label{fig:charge_sharing_explanation}}
\end{figure}

A total input charge of 10 $ke^-$ was injected with test pulses to a cluster of two pixels as seen in Figure \ref{fig:charge_sharing_explanation}. More specifically, 100 consecutive test pulses were distributed across various coordinates within the matrix. Similar to situations involving charge diffusion in thick sensors (in this example from right pixel P2 to the left pixel P1), we distribute the charge in such a way that a portion, denoted as Q\_P1, was directed to pixel P1, while the remaining charge (10 $ke^-$ - Q\_P1) was collected by its neighboring pixel, P2. The lower energy threshold VTH\_L was set at 1 $ke^-$, while the higher threshold VTH\_H could be configured to an energy lower than 10 $ke^-$. An ideal pixel detector should possess the capability to accurately reconstruct both the energy and spatial distribution of each incoming hit, without encountering issues of hit loss or multiple counts, under any setting of VTH\_H lower than 10 $ke^-$. The ideal response of the counters associated with the higher energy threshold for both pixels is depicted in Figure \ref{fig:charge_sharing_explanation}-(c), and Figure \ref{fig:charge_sharing_explanation}-(d) illustrates the total hits registered by both pixels, showcasing the absence of hit loss or multiple counts.

\begin{figure}[htbp]
\centering
\includegraphics[width=1\textwidth]{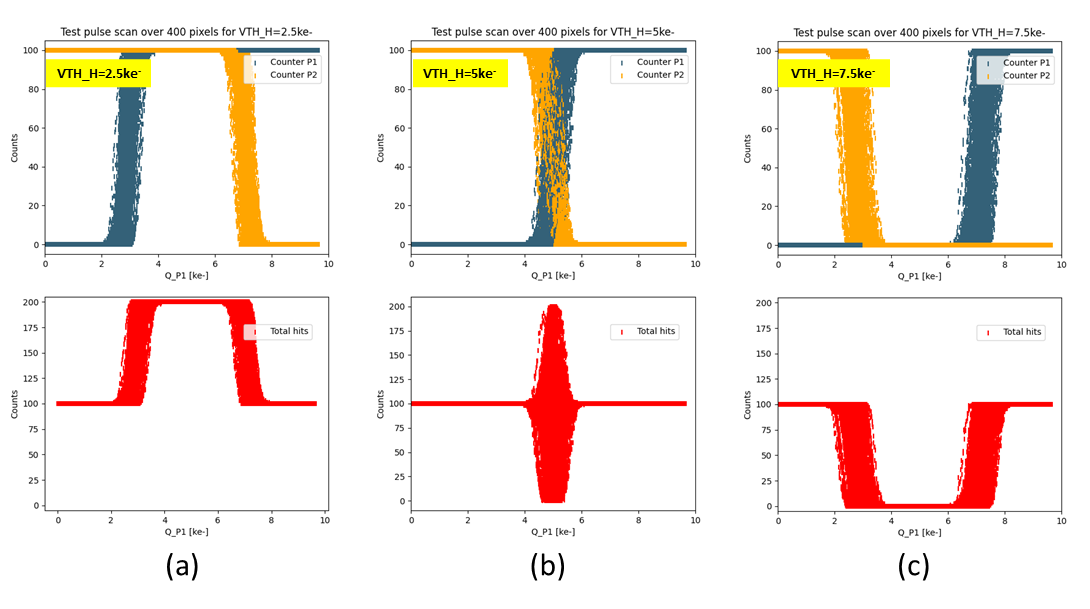}
\caption{Charge sharing effects between pixels when the chip is configured in \textit{fine pitch mode} and \textit{single pixel mode} depending on the energy threshold level: 2.5 $ke^-$ (a), 5 $ke^-$ (b), and 7.5 $ke^-$ (c). In (a), both pixels record counts in the 2.5 $ke^-$-7.5 $ke^-$ range. The multiple registration of hits near the edges or corners between pixels results in poor imaging. In (c), neither of the pixels records counts in the 2.5 $ke^-$-7.5 $ke^-$ range, indicating that hits occurring near the pixel edges or corners go undetected. (b) offers a favorable compromise for accurately reconstructing the energy and spatial position of incoming hits but may not be replicable in the case of a polychromatic input beam.      \label{fig:charge_sharing_fpm}}
\end{figure}

\begin{figure}[htbp]
\centering
\includegraphics[width=1\textwidth]{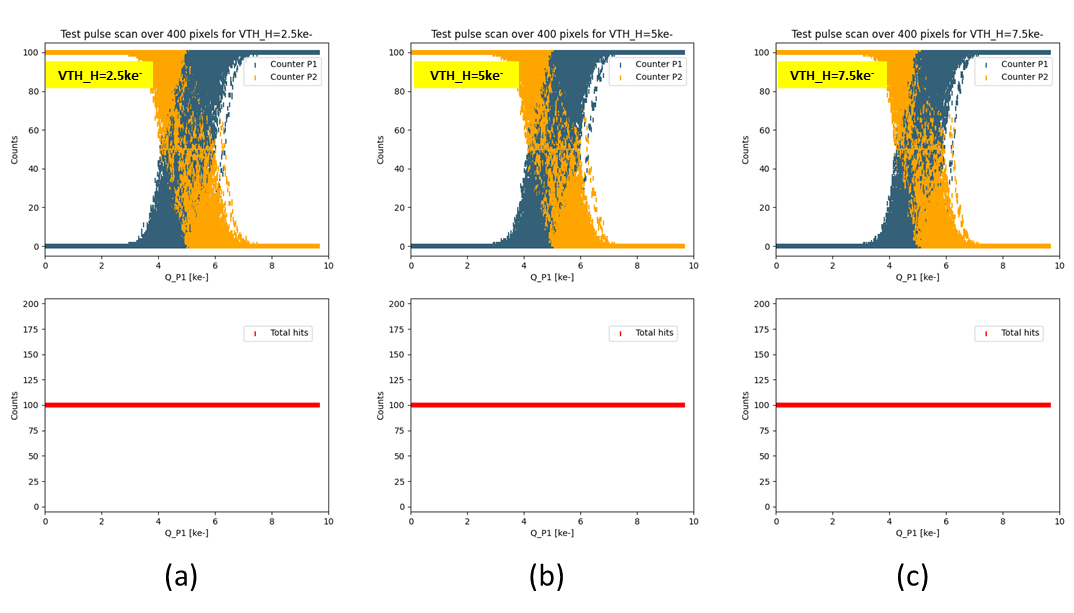}
\caption{Charge sharing effects between pixels when the chip is configured in \textit{fine pitch mode} and \textit{charge summing mode} depending on the energy threshold level: 2.5 $ke^-$ (a), 5 $ke^-$ (b), and 7.5 $ke^-$ (c). Please note that, irrespective of the settings of the higher energy threshold, the s-curves are centered at 5 $ke^-$, showcasing the analog summing capability in \textit{charge summing mode}. Furthermore, each hit is precisely counted once and only once, with no occurrences of undetected hits.       \label{fig:charge_sharing_csm}}
\end{figure}

Figure \ref{fig:charge_sharing_fpm} shows the measured response of the pixels in \textit{fine pitch mode} without charge sharing correction and after threshold equalization. The lower energy threshold was set at 1 $ke^-$, while the higher energy threshold is set at 2.5 $ke^-$ (a), 5 $ke^-$ (b), and 7.5 $ke^-$ (c). The upper plots depict the hits recorded by the higher energy bin for both pixels, P1 (blue) and P2 (orange), while the cumulative hits are presented in the bottom plots (red). These plots illustrate the consequences of charge sharing between pixels, which vary depending on the higher energy threshold level. Setting the higher energy threshold lower than half of the input energy (VTH\_H = 2.5 $ke^-$) leads to the registration of hits near the edges or corners between pixels multiple times. This results in blurred imaging due to charge sharing (and further with fluorescence photons not considered in this setup). Conversely, when employing the higher energy threshold close to the input peak energy (VTH\_H = 7.5 $ke^-$), the hits sharing charge between pixels go undetected. In this scenario, pixels appear to be smaller than their actual size. To mitigate these effects, setting the higher energy threshold at half the input energy may provide a reasonable compromise. However, it's important to note that such effects cannot be controlled in the case of a polychromatic input beam, necessitating on-pixel charge sharing and fluorescence correction mechanisms.
Figure \ref{fig:charge_sharing_csm} illustrates the behaviour of the pixels when configuring the chip with charge sharing correction. In this setup, regardless of the chosen higher energy threshold level, which is set lower than the input energy, each hit is accurately counted once and only once, with no instances of undetected hits. The implementation of this algorithm ensures precise energy and spatial reconstruction for every event, ultimately leading in excellent spectral fidelity.

\begin{table}[htbp]
\centering
\small

\begin{tabular}{|p{3.5cm}|p{1cm} p{1cm}|p{1cm} p{1cm}|p{1cm} p{1cm}|p{1cm} p{1cm}|}
\hline
\multicolumn{1}{|c|}{} & \multicolumn{4}{c|}{\textbf{Fine Pitch Mode (75 µm pixel)}}& \multicolumn{4}{c|}{\textbf{Spectroscopic Mode (150 µm pixel)}}\\
\multicolumn{1}{|c|}{} & \multicolumn{2}{c|}{SPM} & \multicolumn{2}{c|}{CSM} & \multicolumn{2}{c|}{SPM} & \multicolumn{2}{c|}{CSM} \\

 & \textit{UFM} & \textit{LNM} & \textit{UFM} & \textit{LNM} & \textit{UFM} & \textit{LNM} & \textit{UFM} & \textit{LNM} \\
\hline
Pixel gain (${mV/ke^{-}}$) & 10.3$^{*}$ & 19.8$^{*}$ & 10.6$^{*}$ & 18$^{*}$ & 10.4$^{*}$ & 20.1$^{*}$ & 9.4$^{*}$ & 19.4$^{*}$ \\
\hline
Gain variation (${\%}$) & < 3 & < 2 & < 3 & < 2 & < 3 & < 2 & < 3 & < 2 \\
\hline
ENC ($e^{-}$) for bare chip & 116$^{*}$ & 77$^{*}$ & 178$^{*}$ & 117$^{*}$ & 107$^{*}$ & 68$^{*}$ & 172$^{*}$ & 105$^{*}$ \\
\hline
Dynamic range ${(ke^-)}$ & > 25$^{**}$ & > 25$^{**}$ & 42 & 28 & > 25$^{**}$ & > 25$^{**}$ & 42 & 28 \\
\hline
Threshold dispersion before equalization ($e^{-}$) & 953 & 524 & 1430 & 735 & 1185 & 570 & 1300 & 770 \\
\hline
Threshold dispersion after equalization ($e^{-}$) & 77 & 40 & 150 & 95 & 81 & 39 & 125 & 83 \\
\hline
Minimum detectable charge ($e^{-}$) & 830 & 520 & 1400 & 900 & 800 & 475 & 1275 & 800 \\
\hline
Full chip static power consumption (${W/cm^{-2}}$) & 0.46 & 0.46 & 0.47 & 0.47 & 0.31 & 0.31 & 0.36 & 0.36 \\
\hline

\end{tabular}
\caption{Electrical measurements of Medipix4's front-end in the different mode of operations. The characterizations were conducted on a bare readout chip using electrical test pulses and assuming a 5 fF test pulse capacitance ($^{*}$$\pm 30\%$ accurate and $^{**}$limiting the input charge to 25${ke^-}$). \label{tab:i}}
\end{table}

A full summary of the electrical measurements in the eight different mode of configurations is provided in Table \ref{tab:i}. As a key figure, in \textit{ultra-fast mode}, the linear dynamic range of the front-end extends to 42 ${ke^-}$, which is equivalent to 185 keV when employing a CdTe sensor. This signifies more than 50\% improvement compared to Medipix3RX. In \textit{low noise mode}, the readout noise remains comparable to that of its predecessor. Importantly, these crucial front-end achievements are realized while maintaining the full chip static power consumption below 0.5 $W/cm^2$.

%% file: Summary.tex
\section{Summary and future plans}

In this work, we have presented the Medipix4 hybrid pixel detector readout ASIC, describing its architecture and implementation along with the first electrical measurements. The charge sharing correction algorithm works as expected from the simulation. The front-end of the Medipix4 chip can be configured in two different analog modes: \textit{low noise mode} (LNM) and \textit{ultra-fast mode} (UFM). In LNM, the measured electronic noise is 72 $e^-$ r.m.s. using fine pitch configuration without charge sharing correction. In UFM, we expect a linear range up to 42 $ke^-$, showing 50\% improvement compared with Medipix3RX. These performance enhancements compared to its predecessor are achieved at the expense of a slightly larger spatial resolution. The latter should not be a problem since studies have indicated that the optimal pixel pitch when using high-Z materials should be slightly larger than the Medipix3RX pixel in order to account for a larger fraction of fluorescence photons. Detailed measurements of this readout coupled with a sensor will be published when they become available.